\documentclass[proceedings, preprint]{rmaa}

\usepackage{paralist}

\usepackage{psfrag,color}

\SetYear{2007}
\SetConfTitle{Mem\'orias de la Reuni\'on Regional Latino Americana de la UAI}

\title{The Star Formation Histories of galaxies: A tour through the STARLIGHT-SDSS database}

\author{
  R. Cid Fernandes,\altaffilmark{1} 
  W. Schoenell,\altaffilmark{1} 
  J. M. Gomes,\altaffilmark{1} 
  N  V. Asari,\altaffilmark{1} 
  M. Schlickmann,\altaffilmark{1} 
  A. Mateus,\altaffilmark{2}
  G. Stasi\'nska,\altaffilmark{3}
  L. Sodr\'e,\altaffilmark{4}
  and 
  J. P. Torres-Papaqui\altaffilmark{5}
  (the SEAGal collaboration)}

\altaffiltext{1}{Departamento de F\'{\i}sica - CFM, Universidade
    Federal de Santa Catarina, Florian\'opolis, SC, Brazil.}
\altaffiltext{2}{Laboratoire d'Astrophysique de Marseille, CNRS
    UMR6110, Traverse du Siphon, 13012 Marseille, France}
\altaffiltext{3}{LUTH, Observatoire de Paris, CNRS, Universit\'e Paris
      Diderot; Place Jules Janssen 92190 Meudon, France}
\altaffiltext{4}{Instituto de Astronomia, Geof{\'i}sica e Ci\^encias
    Atmosf\'ericas, Universidade de S\~ao Paulo, Brazil}
\altaffiltext{5}{Instituto Nacional de Astrof\'{\i}sica, \'Optica y
   Electr\'onica, Puebla, M\'exico}

\shortauthor{Cid Fernandes \etal}
\shorttitle{RevMexAA(SC) Demo Document}

\listofauthors{R. Cid Fernandes, W. Schoenell, J. M. Gomes, N. V. Asari, M. Schlickmann, A. Mateus, G. Stasi\'nska, L. Sodr\'e, J. P. Torres-Papaqui}

\indexauthor{Cid Fernandes, R.}
\indexauthor{Schoenell, W.}
\indexauthor{Gomes, J. M.}
\indexauthor{Asari, N. V.}
\indexauthor{Schlickmann, M.}
\indexauthor{Mateus, A.}
\indexauthor{Stasi\'nska, G.}
\indexauthor{Sodr\'e, L.}
\indexauthor{Torres-Papaqui, J. P.}

\abstract{Retrieving the Star Formation History (SFH) of a galaxy out
of its integrated spectrum is the central goal of stellar population
synthesis. Recent advances in evolutionary synthesis models have given
new breath to this old field of research. Modern spectral synthesis
techniques incorporating these advances now allow the fitting of
galaxy spectra on an \AA-by-\AA\ basis. These detailed fits are useful
for a number of studies, like emission line, stellar kinematics, and
specially galaxy evolution. Applications of this {\em semi-empirical}
approach to mega data sets are teaching us a lot about the lives of
galaxies. The STARLIGHT spectral synthesis code is one of the tools
which allows one to harness this favorable combination of
plentifulness of data and models. To illustrate this, we show how SFHs
vary across classical emission line diagnostic diagrams. Systematic
trends are present along both the star-forming and active-galaxy
sequences.  We also briefly describe experiments with new versions of
evolutionary synthesis models. Last but not least, we announce the
public availability of both STARLIGHT and a database of detailed
spectral fits and related products for over half a million galaxies
from the SDSS.  This facility allows more physically inspired
explorations of the parameter space than is possible in terms of raw
observed properties, offering new ways to navigate through the realm
of galaxies.}

\resumen{Recuperar la Historia de Formaci\'on Estelar (SFH) de una
galaxia a partir de su espectro integrado es el objetivo central de la
s\'{\i}ntesis de poblaci\'on estelar. Recientes avances en modelos de
s\'intesis evolutiva han dado un nuevo aliento a este viejo campo de
investigaci\'on. Modernas t\'ecnicas de s\'{\i}ntesis espectral
incorporando estos avances hoy permiten el ajuste del espectro de una
galaxia \AA-por-\AA. Estos ajustes detallados son \'utiles para un
n\'umero de estudios, como l\'{\i}neas de emisi\'on, cinem\'atica
estelar, y, por supuesto, evoluci\'on de gal\'axias. Aplicaciones de
esta abordaje {\em semi-emp\'{\i}rica} a mega bases de datos nos
est\'a ense\~nando mucho acerca de la vida de galaxias. El c\'odigo de
s\'{\i}ntesis espectral STARLIGHT es una de las herramientas la cual
permite explotar esa combinaci\'on favorable de datos y modelos. Como
una ilustraci\'on, mostramos como la SFHs var\'ian para galaxias en
distintas regiones de un diagrama de diagn\'ostico cl\'asico. Se
detectan fuertes tendencias sistem\'aticas a lo largo de las
secuencias de formaci\'on estelar y galaxias activas.  Experimentos
con nuevas versiones de modelos de s\'{\i}ntesis son brevemente
descritos.  Por fin, anunciamos la disponibilidad p\'ublica tanto de
STARLIGHT como una base de datos con ajustes espectrales detallados y
productos relacionados para m\'as de medio mill\'on de galaxias del
SDSS. Esta herramienta permite exploraciones f\'{\i}sicamente
inspiradas del espacio de par\'ametros, proporcionando nuevas maneras
de navegar por el reino de galaxias. }

\addkeyword{H~II regions}
\addkeyword{ISM: Jets and outflows}
\addkeyword{Stars: Pre-main sequence}
\addkeyword{Stars: Mass loss}

\begin{document}
\maketitle

\section{Introduction}
\label{sec:intro}

How do galaxies assemble their stars over time, i.e, how do they
evolve?  This broad question lies at the heart of a large number of
astrophysical frontier-problems, from the internal physics of galaxies
to cosmological issues. Theoreticians try to answer this question
plugging in as much physics as they can in their models (see Abadi's
talk), while many observers tackle it using the expanding Universe as
a time-machine, examining how galaxy properties change with redshift
(e.g., Aretxaga's talk). This contribution deals with what can be
learnt from a third and independent method: Uncovering the fossil
record of evolution from its imprints on galaxy spectra.  This {\em
semi-empirical} approach has become highly attractive in the past few
years, given the avalanche of data from cosmologically shallow
surveys, and the enormous progress in our ability to fit galaxy
spectra on a pixel-by-pixel basis using state-of-the-art evolutionary
synthesis models as those reviewed in Bruzual's and Coelho's
contributions.

Taking full advantage of this favorable combination of abundance of
data and models requires spectral synthesis tools to extract
information about age ($t$), metallicity ($Z$) and the detailed star
formation history (SFH) encoded in observed spectra. There are now
several such tools around, differing in both technical and
astrophysical aspects (eg, Mateu's talk).  Some account for extinction
and/or kinematics, others don't. Some impose simple $Z$-$t$ relations
or use a fixed $Z$, while others treat $Z$ and $t$ independently. Some
model indices, others the full spectrum. Some fit the SFH in a
non-parametric fashion, while others compare the data to a library of
precomputed models. Some prefer to compress the input data, others the
output parameters. The list goes on and on\ldots We skip the
impossible task of covering all this ground by referring the reader to
two reviews by our group (astroph/071899 and 071902), dedicated to
basic aspects of spectral synthesis and the recent literature in the
field.

These few pages focus on results obtained with our STARLIGHT synthesis
code. Rather than self-advertisement, this is done with the specific
purpose of illustrating the sort of science doable with this code,
specially when applied to mega data sets. The motivation for this is
that both STARLIGHT and products of its application to $\sim$ all SDSS
galaxies are now available for public use. In {\em
www.starlight.ufsc.br} the reader will find the code itself,
\AA-by-\AA\ fits for 573141 galaxies from the SDSS DR5 and a long and
diversified list of derived properties (from stellar masses to
emission line fluxes). This database is about to $\sim$ double with
694135 DR6 spectra fitted with new evolutionary synthesis
ingredients.

We start with a quick introduction to STARLIGHT, its deliverables and
our VO-like database (\S\ref{sec:STARLIGHT}), including a first ever
comparison of results obtained with DR5 data modeled with the standard
Bruzual \& Charlot (2003) models with DR6 data fitted with newer
models. As an invitation to our database, we illustrate the power of
our detailed spectral synthesis approach by showing how SFHs vary as a
function of location on classical emission line diagnostic diagrams
(\S\ref{sec:SFH_on_BPT}).

\section{The STARLIGHT/SEAGal Project}
\label{sec:STARLIGHT}

\subsection{The code}

STARLIGHT (Cid Fernandes \etal\ 2005)
 combines $N_\star$
spectra from a user-defined base of individual populations in search
of linear combinations which match an input observed spectrum.  The
fitted coefficients define an $N_\star$-dimensional {\em population
vector} (light fractions at a reference $\lambda$). For SFH studies it
is useful to use a base of instantaneous bursts of different $t$'s and
$Z$'s, but anything else can be used. Kinematics is allowed for, as is
reddening (according to any law). Papers by our Semi Empirical
Analysis of Galaxies (SEAGal) collaboration discuss the code and the
results of its application to $\sim$ 0.5M galaxies from the SDSS. For
an in depth description of the code, its pros and cons and possible
uses a detailed 45-pages long manual is available.

STARLIGHT itself outputs: 

\begin{asparaitem}
\item The full synthetic spectrum $M_\lambda$.
\item The light-fraction population vector $\vec{x}$.
\item A mass-fraction population vector $\vec{\mu}$ (only meaningful
for properly defined bases).
\item Stellar velocity dispersion ($\sigma_\star$) and  shift
($v_\star$).
\item Stellar extinction ($A_V^\star$).
\end{asparaitem}

This is already enough for those interested in, say, stellar
kinematics, or if all you need is a decent stellar template to
subtract from your data to aid emission line work. In neither case you
would care about $\vec{x}$ nor $\vec{\mu}$. For galaxy evolution work,
on the other hand, it is precisely the population vector which
matters, since this is where the SFH information is. It can be handled
in numerous ways to produce things like time dependent star formation
rates, SFR(t).

\subsection{The database: www.starlight.ufsc.br}

\begin{figure}[!t]
  \includegraphics[width=\columnwidth]{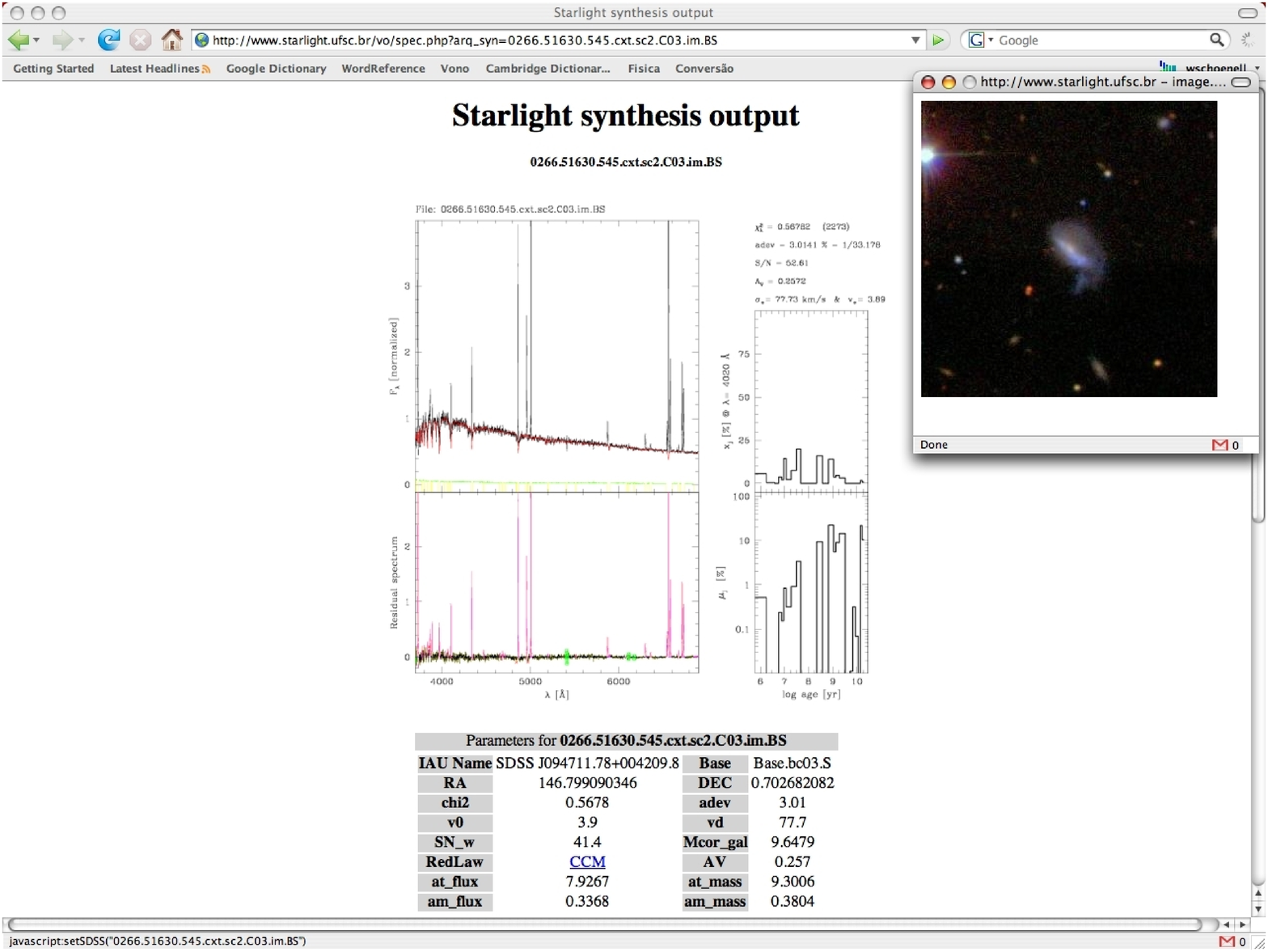}
  \includegraphics[width=\columnwidth]{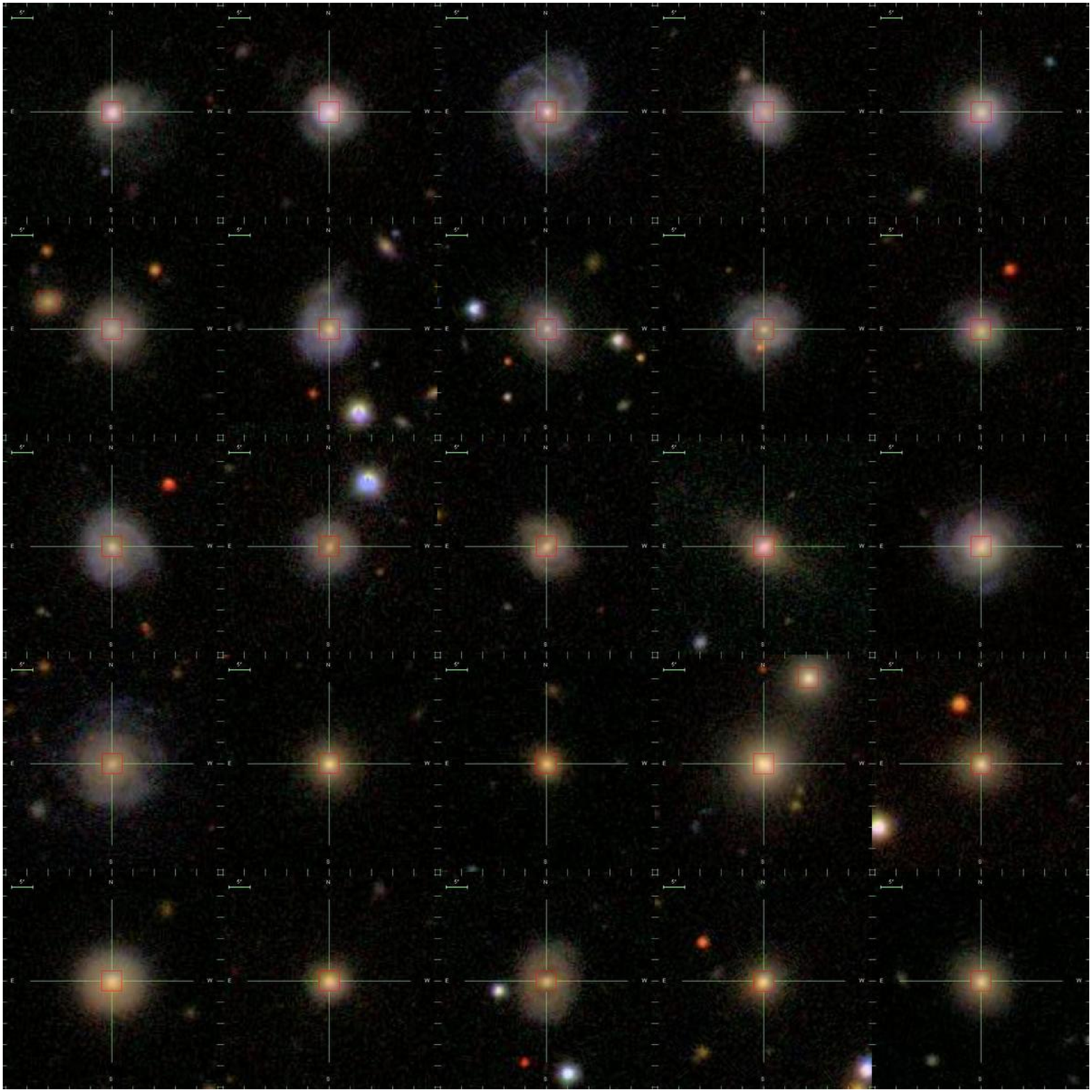}%
  \hspace*{\columnsep}%
  \caption{{\em Top:} Screen-shot showing an example database query.
{\em Bottom:} Galaxies of same stellar mass ($10^{11.00\pm 0.05}
M_\odot$) and redshift ($0.07\pm 0.01$), low inclination ($b/a \ge
0.9$) sorted according to the STARLIGHT-derived (light-weighted) mean
stellar age: $\overline{\log t} [\rm yr] = $8.3 to 10.3, increasing to
the right and from the top.}
  \label{fig:ScreenShotWWW}
\end{figure}

In addition to the data listed above, our STARLIGHT-SDSS database
contains a series of other products:
 
\begin{asparaitem}
\item Emission line properties (fluxes, equivalent widths, widths,
S/N) for many transitions, measured from the starlight-subtracted
spectrum.
\item Stellar masses ($M_\star$), and mass-to-light ratios.
\item Summaries of the population vector, like mean stellar ages and
metallicities.
\item Basic SDSS data, like coordinates, redshifts, magnitudes, sizes
and images.
\end{asparaitem}

Fig.\ \ref{fig:ScreenShotWWW} shows a screen-shot of one object of a
list produced with a SQL selection. The page contains the spectrum,
the STARLIGHT fit and all that is known about the galaxy, as well as
its SDSS thumbnail.\footnote{The whole interface was developed and is
maintained by an undergrad student (W. Schoenell)!}  A particularly
attractive feature is that one may search and select by any
combination of {\em physical} and observed properties. This allows one
to play all sorts of games.  The bottom panel in Fig.\
\ref{fig:ScreenShotWWW} shows an example. We have selected galaxies
within 0.05 dex of $M_\star = 10^{11} M_\odot$, $z = 0.07 \pm 0.01$
and produced a mosaic of 25 images sorted according to the mean
stellar age. The progression from late to early type morphologies is
evident to the eye.

\subsection{Elementary (but useful) warnings}

It is never too much to emphasize the following obvious, but often
overlooked, points: (1) The inversion from an observed spectrum to a
set of parameters is far more complex and degenerate than one would
hope. Yet, as long as one does not fall into the temptation of
embracing over-detailed descriptions of SFHs, all modern synthesis
methods seem to converge to the same overall result. Furthermore,
averaging results over hundreds of galaxies greatly alleviates
uncertainties in individual SFHs (Panter \etal\ 2007). (2) To go from
an observed galaxy spectrum to a population vector and its associated
SFH one {\em must} chose one among {\em many} possible
bases. Evolutionary synthesis models themselves are undergoing
constant revisions, approximately at the same rate of that of
improvements in their two main ingredients: stellar tracks and
spectral libraries---this very same volume contains 2 such updates!
(3) Finally, observational data are never problem-free. We have chosen
to fit anything spectroscopically classified as a galaxy in the SDSS!
This includes quite a bit of garbage, which must be filtered out with
appropriate quality flags.

\subsection{New bases +  new data = more work}

\begin{figure*}[!t]
  \includegraphics[width=\textwidth]{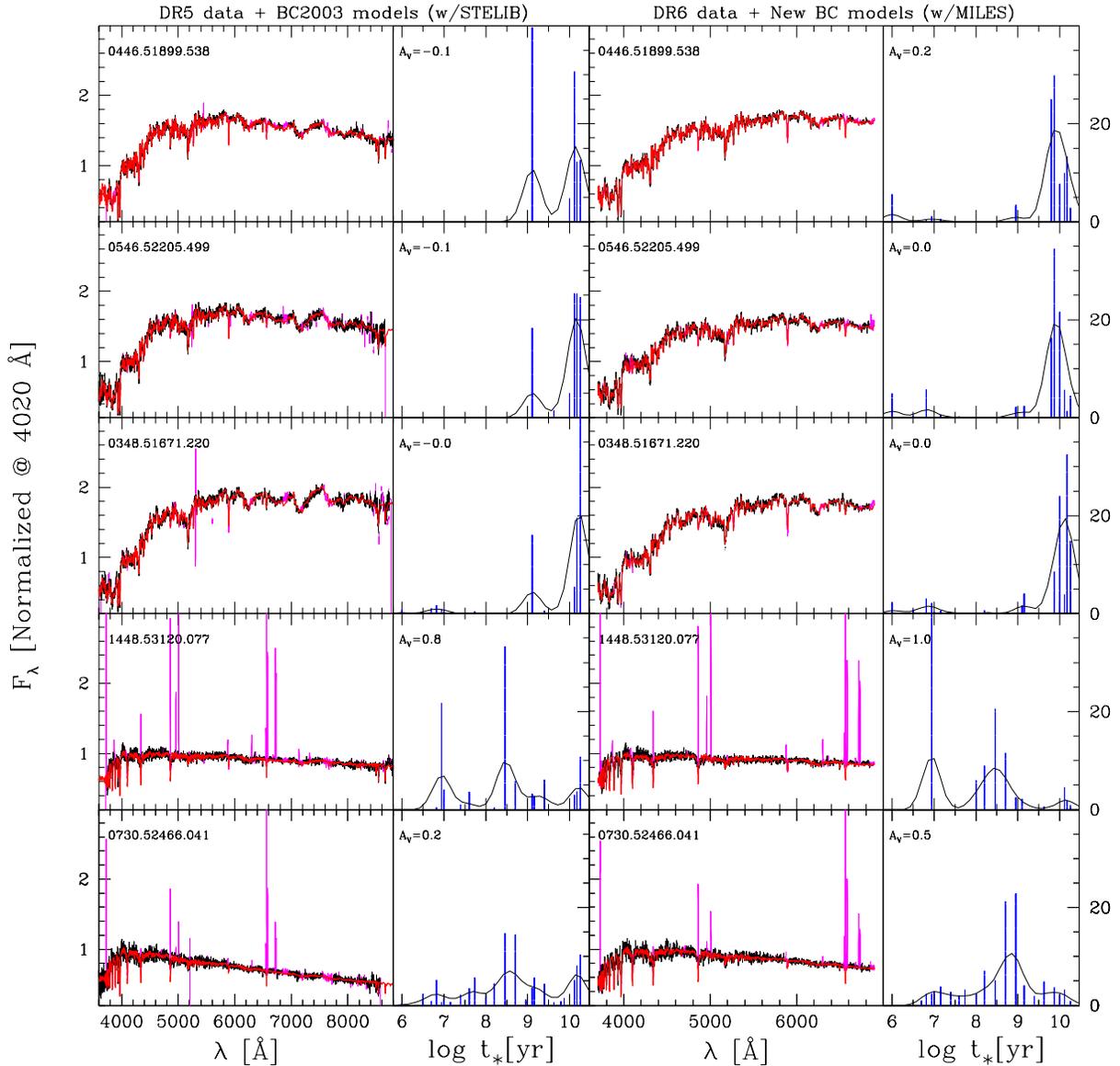}%
  \hspace*{\columnsep}%
  \caption{Fits of the same five SDSS galaxies, but using DR5 +
Bruzual \& Charlot (2003) models with the STELIB library (left) or the
DR6 version with a newer version of the models with MILES
(right). Black and red are the observed and synthetic spectra, with
masked regions plotted in magenta. The histograms show the \% {\em
light fractions} at 4020 \AA\ from the STARLIGHT decomposition onto a
common set of 25 ages $\times$ 6 metallicities. Solid lines are 0.5
dex smoothed versions of our population vector. (Given the 4 dex
dynamic range in $t_\star$, smoothing by 0.5 dex is qualitatively
equivalent to reducing the base to 8 ages, perhaps still a bit
over-optimistic.)  Despite the broad-brush agreement, there are
substantial differences both in the input data and in the fitted
SFHs. Notice also the different spectral ranges.  }
 \label{fig:Comparison}
\end{figure*}

There are news about points 2 and 3 in the above list. First, we point
out that all published results by the SEAGal collaboration rely on the
2003 version of the Bruzual \& Charlot models with the STELIB library
and Padova 1994 tracks applied to over 0.5M SDSS DR5 spectra. But
there is a lot more coming.

As discussed by G. Bruzual elsewhere in this volume, there are newer
versions of these models, differing both in the treatment of late
evolutionary phases and in the spectral libraries. We have carried out
a series of experiments with some of these new models, and found that:
(a) Not surprisingly, the new evolutionary tracks have negligible
impact on our optical spectral synthesis (they only affect the
IR). (b) Fits with the MILES library (S{\'a}nchez-Bl{\'a}zquez et al.\
2006) provide substantial improvement with respect to those obtained
with STELIB, particularly for old systems. Besides smaller spectral
residuals, these new models correct some pathologies in the derived
physical properties of passive/elliptical galaxies, like weird-looking
distributions of sources in the $\overline{t}$-$\overline{Z}$ plane
and (slightly) negative values of $A_V^\star$. Things are thus
improving on the modelling front.

There are also news about the data: The SDSS reduction pipeline
changed, and DR6 spectra are different in shape and amplitude from the
DR5 ones.

We have just finished fitting 694135 DR6 spectra with the new
MILES-based models, and will soon incorporate these data to our
database.  Fig.\ \ref{fig:Comparison} presents a preliminary
comparison of old and new fits to three passive and two SF
galaxies. Some of the changes mentioned above are visible in these
examples. The similarities and differences revealed by practical tests
like this provide useful feedback to model makers, and portray a
summary of the degree of maturity in the field. While these new models
represent a perceptible improvement, before jumping to conclusions one
must realize that there are many others available, some of which
address known deficiencies in standard models (like
$\alpha$/Fe). Examining all possibilities and digesting the results
will take time and patience. It will be particularly interesting to
investigate the effects of different stellar evolution tracks, even if
that requires extending the spectral domain beyond the optical
(e.g., Eminian \etal\ 2008).

\section{Star Formation Histories across the BPT diagram}

\label{sec:SFH_on_BPT}

\begin{figure*}[!t]
  \includegraphics[width=\textwidth]{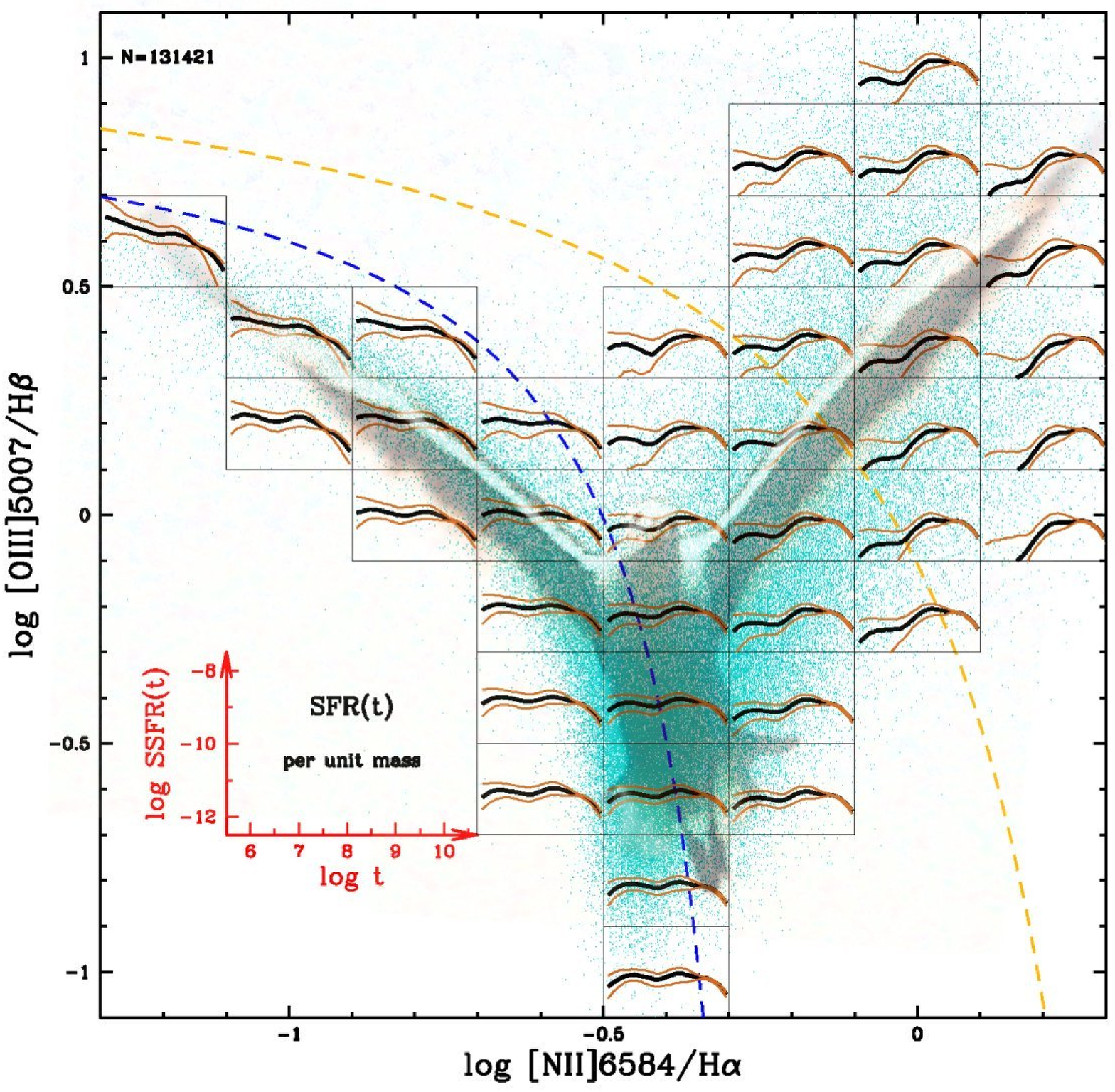}%
  \hspace*{\columnsep}%
  \caption{BPT/``seagull'' diagram for 131421 SDSS galaxies with
superimposed SFHs as derived by STARLIGHT. The dashed lines show the
SEAGal (Stasi\'nska \etal\ 2006, in blue) and Kewley \etal\ (2006,
orange) AGN/SF dividing lines. Each box shows the $t$-by-$t$ median
(thick black line), 16 and 84 percentiles (thinner brown lines) of the
time dependent SFR per unit mass for galaxies within the box. The
inset shows the scale (units are yr$^{-1}$ for SSFR and yr for
lookback time $t$). Note the systematic progression of the ratio of
past to present star formation as one moves down the SF (left) wing,
and from the Seyfert to the LINER branches (above and below the
seagull's right wing, respectively).}
  \label{fig:BPT}
\end{figure*}

Emission lines reflect the physical conditions of the ISM and the
nature of the ionizing source. Since Baldwin, Phillips \& Terlevich
(1981), diagrams involving pairs of line ratios are used to diagnose
galaxies as star-forming (SF) or active (Seyfert or LINER), and to
derive properties like the nebular metallicity ($Z_{neb}$), N/O ratio
and ionization parameter. An interesting question to ask is: How do
the SFHs of galaxies change across such diagnostic diagrams? Given
that the ISM conditions are more a consequence than a cause of galaxy
evolution, this question is kind of an exercise in reverse
engineering.

Fig.\ \ref{fig:BPT} shows the classical
[\ion{O}{iii}]$\lambda$5007/H$\beta$ vs.\
[\ion{N}{ii}]$\lambda$6584/H$\alpha$ BPT diagram. The background in
this plot consists of emission line measurements for 131421 galaxies
from our STARLIGHT-SDSS database). The SF and AGN sequences are
clearly visible, branching out like the wings of a seagull. We have
chopped this seagull into small boxes and computed the statistics of
the $t$-by-$t$ SFH within each box (like in Asari \etal\ 2007, but
including AGN). The thick-black line shows the time evolution of the
median Specific SFR (SSFR), whereas the thiner-brown lines are the 16
and 84 percentiles.

Systematic variations in the evolutionary pattern are seen all across
the diagram. Along the left wing, for instance, all galaxies show
current SF (as expected), but the balance between current and past SFR
changes by over an order of magnitude. That is why galaxies at the top
left, where low-$Z_{neb}$ galaxies live, look much younger than the
metal-rich systems at the bottom of the sequence. Mass increases
along this sequence (the $M_\star$-$Z_{neb}$ relation), so this pattern
is ultimately a consequence of galaxy downsizing.

Strong trends are also evident in the AGN wing. The stellar population
mixture becomes increasingly skewed toward older ages as one moves up
the AGN wing, in good part due to aperture effects. Even concentrating
only on the region far away from the body of the seagull (eg, beyond
the Kewley line, where contamination by off nuclear SF regions is
minimal), one sees trends: Seyfert 2s (upward in the plot) are
systematically younger than LINERs. In most AGN the ongoing SF is but
a fraction of what it has once been, and this fraction tends to zero
as one walks the BPT diagram from the Seyfert to the LINER
territories. Several other things change on the way too: $A_V$ and
line luminosities go down, while $M_\star$, $\overline{Z_\star}$ and
$\alpha$/Fe increase.

\begin{figure}[!t]
  \includegraphics[width=\columnwidth]{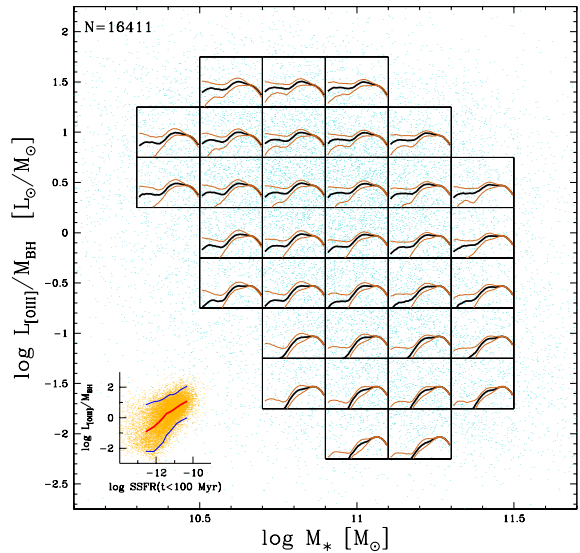}%
  \hspace*{\columnsep}%
  \caption{As Fig.\ \ref{fig:BPT}, but in the ``Eddington ratio'' vs
  galaxy mass plane, and restricted to the 16141 galaxies above
  Kewley's maximal-starburst line. The inset shows the relation between
  the specific SFR over the last $10^8$ yr and a proxy for the
  ``specific black hole accretion rate''.
}
  \label{fig:LO3_MBH}
\end{figure}

To have a closer look at this trend, Fig.\ \ref{fig:LO3_MBH} shows
SFHs in the $L_{\rm [OIII]} / M_\bullet$ vs.\ $M_\star$ plane, where
$M_\bullet$ is the indirectly computed black hole mass (from
$\sigma_\star$). The y-axis discriminates between Seyferts and LINERs
(Kewley \etal\ 2006), and is usually interpreted as a proxy for the
Eddington ratio (Heckman \etal\ 2004).  Only 16411 galaxies ``pure
AGN'' are shown in this plot. Despite the relatively compressed
$M_\star$ scale (AGN live in massive galaxies), it is interesting to
note that for a fixed $L_{\rm [OIII]} / M_\bullet$ the SFHs are
essentially {\em independent} of galaxy mass. On the contrary, for any
given $M_\star$ the ``Specific Black Hole Accretion Rate'' (i.e.,
$\dot{M}_\bullet / M_\bullet$) is obviously linked to the recent SFH:
Black holes are growing more efficiently in galaxies which are forming
stars more efficiently. This is shown more directly in the inset,
which plots the relation between $L_{\rm [OIII]} / M_\bullet$ and the
mean SSFR over the last 100 Myr. 

A few years ago this would be called ``the Starburst-AGN connection''.
To find out whether this fits with negative AGN-feedback scenarios
fashionable nowadays, read our next papers. Or else, do it yourself at
www.starlight.ufsc.br!

\end{document}